\documentstyle[11pt,newpasp,twoside,epsfig]{article}
\index{summary}
\index{instructions}
\index{template}

\newcommand\kms{km\,s$^{-1}$}
\newcommand\halpha{H$\alpha$\,\,}

\def\edcomment#1{\iffalse\marginpar{\raggedright\sl#1\/}\else\relax\fi}
\reversemarginpar

\begin{document}
\title{Hydrodynamical Models of Pulsar Wind Bow-Shock Nebulae}
 \author{Niccol\`o Bucciantini}
\affil{Dipartimento di Astronomia e Scienza dello Spazio, Universit\`a di Firenze, L.go E.Fermi 5, Firenze, Italy}

\begin{abstract}
We present hydrodynamical simulations, using a 2-D two fluid model, of bow shocks in a representative regime for pulsar wind driven bow-shock nebulae. We also investigate the behaviour of a passive toroidal magnetic field wounded around the pulsar velocity direction. Moreover we estimate the opacity of the bow-shock to penetration of ISM neutral hydrogen. Finally we compare these numerical results with those from an analytical model. 
\end{abstract}
\section{Introduction}
Pulsars moving supersonically with respect to the ambient medium are expected to give rise to a bow shock. In the case of interaction with the interstellar medium (ISM) the \halpha  emission from the nebula may be detected. Only four bow shocks have been discovered so far: PSR~1957+20 (Kulkarni \& Hester 1988), PSR~2224+65 (Cordes, Romani \& Lundgren. 1993), PSR~J0437+4715 (Bell 1995), and PSR~0740--28 (Jones, Stapper \& Gaensler 2001). These nebulae may be detected in optical Balmer lines (mostly \halpha) as a signature of a non-radiative shock moving through a partially neutral medium (Chevalier \& Raymond 1980).

A puzzling point is that two out of the four known nebulae show a peculiar shape with a conical tail. Various hypothesis have been put forward to justify the conical tail shape: a peculiar ISM density distribution in the surroundings of the PSR~2224+65; effects like mass loading due to neutral atoms which might penetrate the external layers of shocked material (Bucciantini \& Bandiera 2001, hereafter Paper I) . The nebula near PSR~0740--28, seems to represent an intermediate case, with a ``standard'' head and a conical tail.
The ISM is seen by the pulsar as a plane-parallel flow, likely with a constant density, and lasting long enough to produce a steady-state regime. Moreover the ISM is typically partly neutral and the neutral atoms may have collisional mean free paths comparable with the scale length of the system, then invalidating a purely fluid treatment. However as demonstrated in a previous paper (Paper I), if we suppose that the H atoms can be ionized only via collisions, for a large amount of pulsars the presence of a neutral component in the ISM can be neglected . 

The interaction of the relativistic magnetized pulsar wind with the ISM ionized component is due to the pulsar wind magnetic field compressed on the head of the nebula. The mean free path of particles is typically much smaller than the typical dimension of the nebula: $d_{o}=\sqrt{{\cal L}/4\pi c \rho_{o} V_{o}^{2}}$,where ${\cal L}$ and $V_{o}$ are the pulsar luminosity and velocity, and $\rho_{o}$ the density of the dragged component of ambient medium (Paper I).
\section{Numerical Simulation}
We have used the multi-D code CLAWPACK (LeVeque 1994): a second order Godunov with piecewise linear reconstruction. The problem has an intrinsic cylindrical symmetry which allows us to use a two-dimensional grid. Both the stellar and the external winds are taken with high Mach numbers (hypersonic limit). In fact the pulsar wind has a thermal pressure much lower than its ram pressure and the same holds true for the ISM.
We decided to use a different adiabatic coefficient for the ambient medium (with adiabatic coefficient $5/3$) and the relativistic ($4/3$) fluid coming from the star.
The physical conditions of such a bow-shock guarantee conservation of energy, and little mixing between the two media. However we added a little diffusion to avoid the growth of numerical instabilities.
 {\em Ul} is the unit length we have used. For the simulations we used a grid of 580$\times$400 cells and the star in (420,0). The stellar wind has a velocity of 1000 \kms, a density of 0.04 cm$^{-3}$ and a Mach number about 13 upstream of the termination shock along the bow shock axis. We have performed various simulations using different velocities for the external flow (400, 300, 200, 150 \kms). The time needed to reach the steady state condition is of order 4-5 times the external flow crossing time in the computational box.
\section{Comparison with the analytic model}
The geometry and internal structure of bow-shock nebula resulting from our simulations are shown in Fig.~1. The positions of the termination shock, bow shock and contact discontinuity do not change among all the above mentioned simulations, and also the position of the sonic point remains essentially the same. This is because the solution, in the hypersonic limit, depends only on the ratio between ram pressures of the stellar and ambient medium, but not separately on their densities or velocities. The dashed lines represent two analytic solutions (Wilkin 1996), one scaled to match the contact discontinuity near the head, and the other to match the bow-shock near the head. Fig.~1 shows that the region of unperturbed stellar wind extends in the tail up to a ``spherical shock'', whose position is strictly connected with the value of the thermal pressure in the ambient medium.
The contact discontinuity in the tail (Fig.~1) tends to reach a limiting distance $Z$ from the axis, with radius of the same order of the stagnation point distance.
In order the stellar wind material to expand sideway beyond $Z$, some process is required, which reduces the velocity in the layer or increases its pressure. This can be achieved via processes like mass accretion, or if some further shock develops. 
The distance of the bow-shock from the star given by the analytic model is 0.175 {\em Ul} while the two limiting analytic solutions of Fig.~1 (dashed lines) are at 0.22 and 0.29 {\em Ul}.
The behaviour of the fluid in regions A and B may affect the synchrotron emission from the relativistic electrons and positrons coming from the pulsar and shocked in the termination shock. If particles have relativistic motions their synchrotron emission can be detected only if their pitch angle is close to the observer direction. In the tail the relativistic bulk motion produce a beaming of the emission in the backward direction.

To verify if the two-thin layers model could be taken as a reasonable approximation of the true structure of the nebula, at least as far as the external layer is concerned, and to evaluate how far in the tail it could be used, we have extracted from our simulations the surface density and the tangential velocity in the external layer, between the bow shock and the contact discontinuity.
Fig.~2-left compares the surface density in the external layer, extracted from the simulation, with that evaluated by a two-thin layers model. The analytic curve that best reproduces the numerical points gives an external density of 0.26 cm$^{-3}$ (instead of the true value 0.25 cm$^{-3}$)  and a stagnation point distance of 0.235 {\em Ul}, compatible with the range of the two limit solutions. 

In the same way we have evaluated the tangential velocity in the layer between the bow shock and the contact discontinuity. Fig.~2-right shows the numerical values of tangential velocity in the external layer, normalized to the external wind value, and compares it to the two-thin layers model. The analytic curve that reproduces the points has been obtained using a values of 0.25 {\em Ul} for the distance of the stagnation point, a value in agreement with the two limit solutions. 
\section{Magnetic field and penetration thickness for neutral hydrogen}
Using the numerical hydrodynamic solutions described above we have also evaluated the behaviour of a toroidal magnetic field wounded around the symmetry axis and passively advected by the fluid. We considered the case of a pulsar wind with the energy associated to the ordered magnetic field much smaller than the ram pressure of the relativistic particles, and with the ratio between the two being independent of the direction.  The value of such a field increases in the head, due to the compression in the termination shock, ranging from 7 times its upstream value, suddenly after the shock, up to a factor 30-35 with respect to its upstream value, near the contact discontinuity.
Even if initially the magnetic field's energy is below equipartition, it can reach equipartition if the external wind Mach number is high enough.

As we have demonstrated in a previous paper (Paper I) the penetration thickness $\tau$ for neutral hydrogen in the external layer, depends on ambient density and pulsar velocity. The penetration thickness depends by the processes that could drag the atoms: essentially by charge-exchange and by ionization via the interaction with the protons confined in the external layer. In our previous paper we have evaluated the thickness using the two-thin layers model limited to a region near the axis. Our aim is to verify if such approximation is acceptable and how far in the tail it can be extended. For typical  pulsar velocities, collisional ionization can be neglected, even in non equilibrium cases.
The thickness decreases at higher velocities as expected: in fact it follows the behaviour of the cross section for charge-exchange, which is the dominant process. The thickness, at least for the head, increases slightly: only over a factor 2. This can be quite important, because, if the nebula is thin to the penetration of neutral hydrogen in the head, it can remains thin even in tail.
 
\begin{figure}
\centerline{\epsfig{file=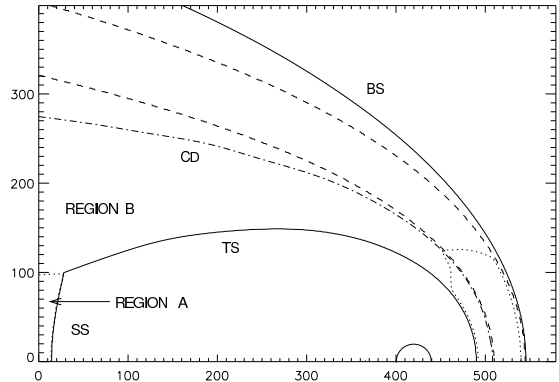,height=5cm} }
\caption{TS = termination shock, SS = spherical shock, CD = contact discontinuity, BS = bow shock: the dashed lines are two analytical solutions, the dotted lines are the sonic surfaces.}
\end{figure}
\begin{figure}
\centerline{\epsfig{file=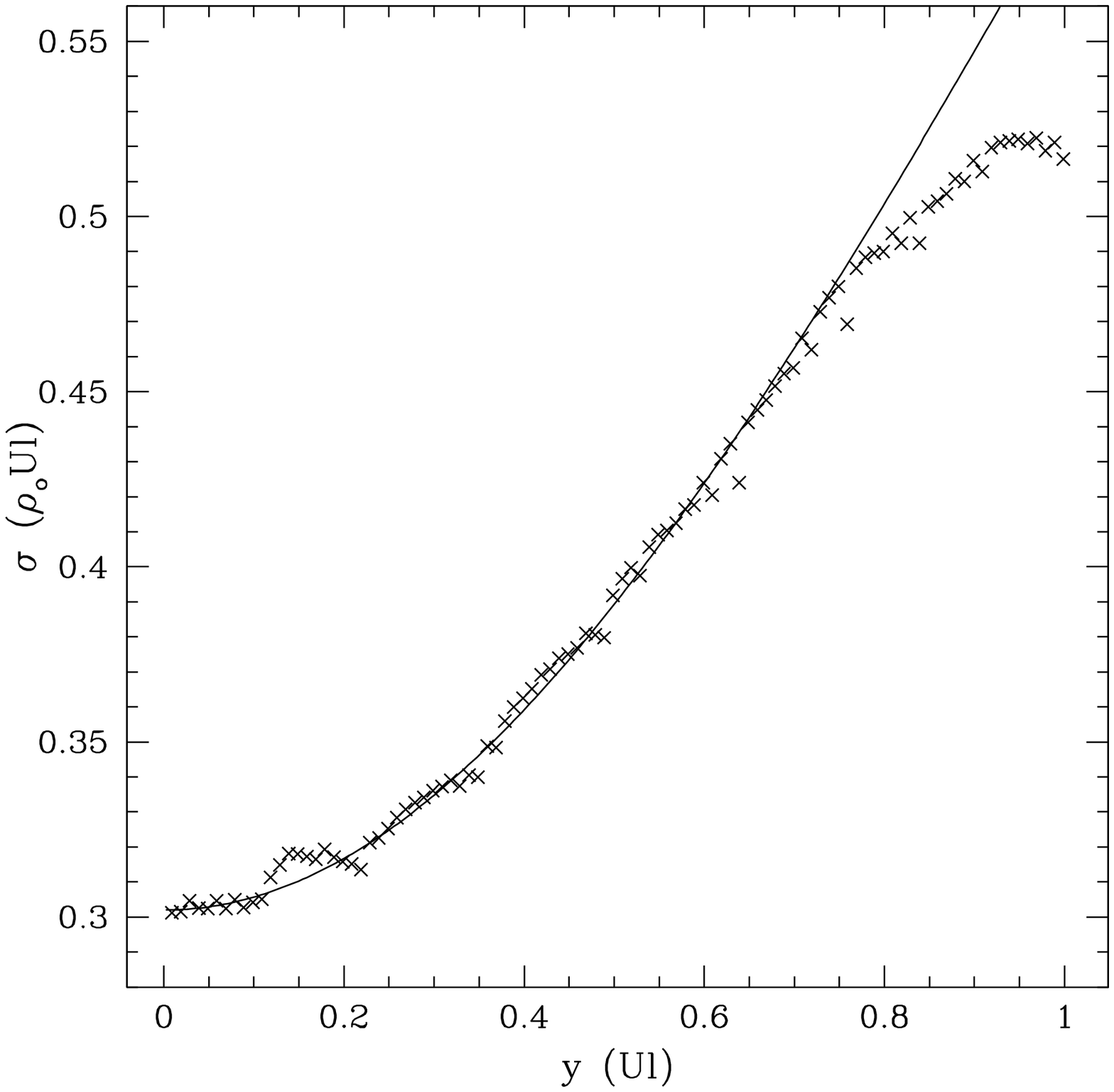,height=5.cm} \hspace{1cm}\epsfig{file=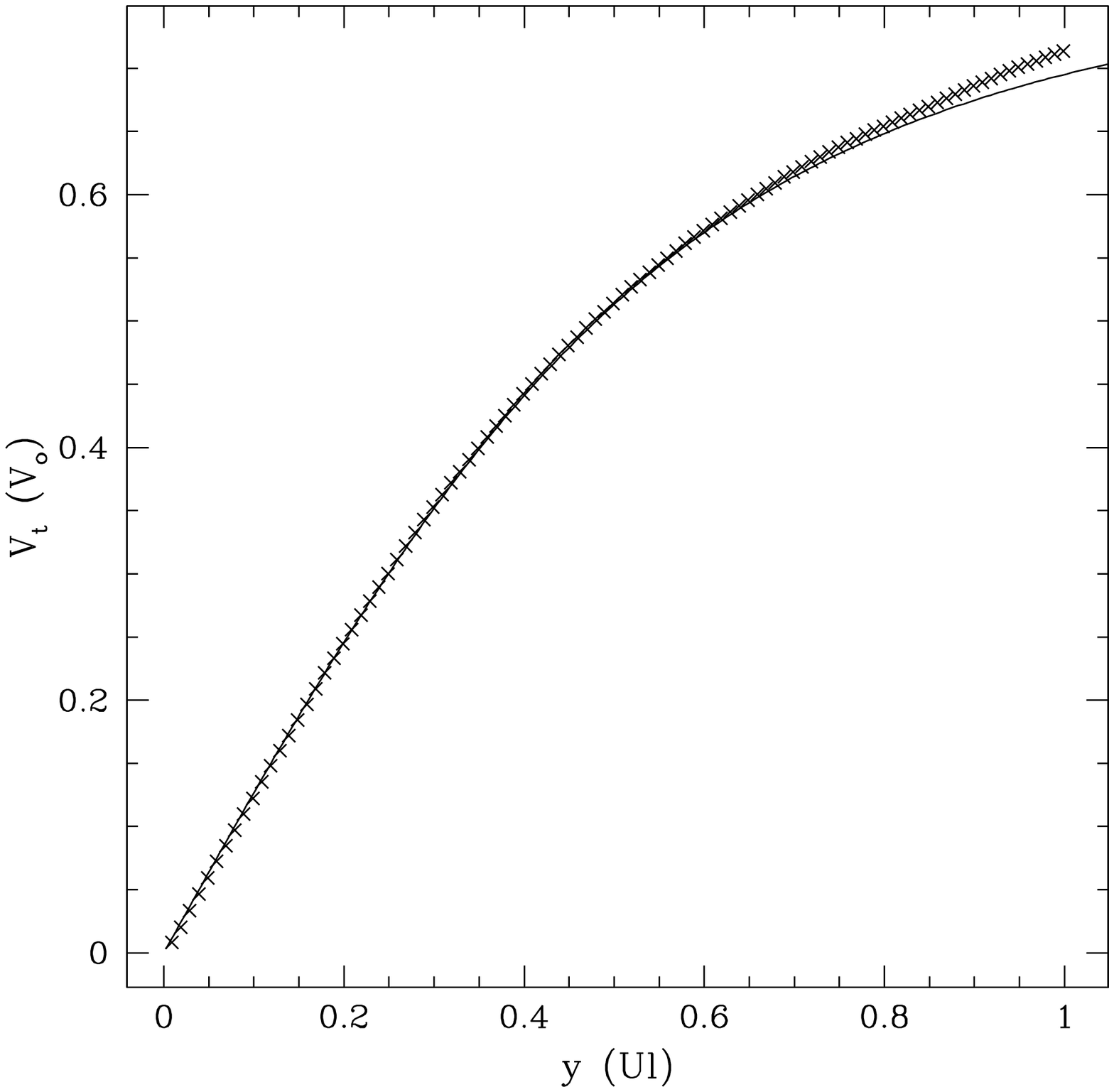,height=5cm}}
\caption{ Left: external layer surface density (crosses) compared with analytic model. Right: external layer tangential velocity (crosses) compared with analytic model. y = distance from the axis.}
\end{figure}


\begin{references}
\reference Bell, J. F., Bailes, M., Manchester, R. N., Weisberg, J. M., \& Lyne, A. G. 1995, \apj, 440, L81
\reference Buciantini, N., \& Bandiera, R. 2001, \aap, 375, 1032
\reference Chevalier, R. A., \& Raymond, J. C. 1978, \apj, 225, L27
\reference Cordes, J. M., Romani, R. W., \& Lundgren, S. C. 1993, \nat, 362, 133
\reference Jones, H., Stappers, B., \& Gaensler, B. 2001, The Messenger, 103, 27
\reference  Kulkarni, S. R., \& Hester, J. J. 1988, \nat, 335, 801
\reference LeVeque, R. J. 1994, Proceedings of the Fifth International Conference on Hyperbolic Problems.
\reference  Wilkin, F. P. 1996, ApJ, \apj, L31
\end{references}
\end{document}